\documentclass[a4paper,fleqn]{cas-sc}\usepackage{lineno}
\usepackage[authoryear]{natbib}
\usepackage[normalem]{ulem}

\def\tsc#1{\csdef{#1}{\textsc{\lowercase{#1}}\xspace}}
\tsc{SJB}
\tsc{JWN}
\tsc{BS}
\tsc{JO}
\tsc{DB}
\begin{document}
\let\WriteBookmarks\relax
\def\floatpagepagefraction{1}
\def\textpagefraction{.001}
\newcommand{\edit}[1]{\textcolor{black}{#1}}

\shorttitle{CS Fluorescence}

\shortauthors{Bromley \textit{et al.}}

%\title [mode = title]{Updated Fluorescence Efficiencies of the CS radical}  

\title[mode = title]{{Updated Ultraviolet Fluorescence Efficiencies of CS:  Evidence for Model Discrepancies in the Enhancement of NUV-Derived CS Abundances in Comets}}

\tnotemark[1]

\tnotetext[1]{This document contains results of a research
   project funded in part by Space Telescope Science Institute (AR 17031). Experiments were funded in part by Europlanet travel grant 22-EPN3-107.}

%\tnotetext[2]{The second title footnote which is a longer text matter
 %  to fill through the whole text width and overflow into
  % another line in the footnotes area of the first page.}

\author[1]{S.J. Bromley}[type=editor,
                        orcid=0000-0003-2110-8152]

\cormark[1]

\ead{sjb0068@auburn.edu}

%\ead[url]{www.cvr.cc, cvr@sayahna.org}

\credit{Conceptualization, Methodology, Software, Investigation, Writing - Original Draft}

\affiliation[1]{organization={Auburn University},
    addressline={Department of Physics}, 
    city={Auburn},
    % citysep={}, % Uncomment if no comma needed between city and postcode
    postcode={AL, 36849}, 
    country={USA}}

\affiliation[2]{organization={Comenius University in Bratislava},
    addressline={Department of Experimental Physics, Faculty of Mathematics, Physics and Informatics},
    city = {Bratislava},
    country = {Slovakia}}
\author[1]{J. Wm. Noonan}[orcid=0000-0003-2152-6987]
\credit{Writing - Original Draft, Conceptualization}
\author[2]{B. Stachov{\'a}}[orcid=0009-0007-2894-3085]
\credit{Investigation, Resources}
\author[2]{J. Orsz{\'a}gh}[orcid=0000-0002-4309-6060]
\credit{Investigation, Resources}
\author[1]{D. Bodewits}[orcid=0000-0002-2668-7248]
\credit{Conceptualization, Writing - Review \& Editing}


\cortext[cor1]{Corresponding author}
%\cortext[cor2]{Principal corresponding author}

%\fntext[fn1]{This is the first author footnote. but is common to third
%  author as well.}
%\fntext[fn2]{Another author footnote, this is a very long footnote and
%  it should be a really long footnote. But this footnote is not yet
%  sufficiently long enough to make two lines of footnote text.}

%\nonumnote{This note has no numbers. In this work we demonstrate $a_b$
%  the formation Y\_1 of a new type of polariton on the interface
%  between a cuprous oxide slab and a polystyrene micro-sphere placed
%  on the slab.
%  }

\begin{abstract}
Observations of carbon monosulfide (CS) have a long history serving as a remote proxy for atomic sulfur, and more broadly, one of the sulfur reservoirs in cometary bodies. {Recently, systematic discrepancies between NUV- and radio-derived CS abundances have been found to exceed a factor of 2 - 5, with NUV-derived abundances appearing enhanced for a wide array of comets. Interpretation of cometary CS emission in the ultraviolet has relied on a murky and ill-documented lineage of calculations whose accuracy can be difficult to assess.} We report new fluorescence efficiencies of the CS radical, utilizing a rovibrational structure with vibrational states up to $v = 8$ and rotational states up to $N = 100$. {The models utilize a new set of band transition rates derived from laboratory electron impact experiments. Benchmark comparisons to IUE observations of C/1979 Y1 (Bradfield) show favorable agreement with the fluorescence models. The present results affirm the accuracy of the historical record of CS abundances derived via ultraviolet CS emission in comets with IUE and HST, but do not explain the consistent enhancement of NUV-derived CS abundances relative to the radio measurements during the same apparitions. Alternative explanations of the factor of 2 - 5 discrepancy between NUV- and radio-derived CS abundances are discussed, as well as possible connections to sulfur reservoirs in protoplanetary disks.} The model code and computed fluorescence efficiencies are made publicly available on the Zenodo service.
\end{abstract}

% \includegraphics{figs/grabs.pdf}

%\item Research highlights item 1
%\item Research highlights item 2
%\item Research highlights item 3

% Keywords

\maketitle

\section{Introduction}
The astrochemical pathways leading to the formation of sulfur and sulfur-bearing species remain poorly understood. The abundance of atomic sulfur in the interstellar medium is nearly 1000 times larger than the abundances measured in molecular clouds and protoplanetary disks \citep{Tieftrunk1994,LeGal2019,kama2019abundant}. 
One of the few sulfur-bearing molecules that has been detected in molecular clouds, protoplanetary disks, and our own solar system's remnant planetesimals, like comets, is carbon monosulfide, or CS {\citep[and sources therein]{Tieftrunk1994,LeGal2019,Teague2018,jackson1982production,roth2021leveraging,biver2022}}. In astrophysical context the molecule is an important tracer of the C/O ratio within protoplanetary disks when observed in tandem with SO in the {radio}~\citep{Keyte2023azimuthal}. While CS is commonplace in the planetary formation process, establishing its role and significance in the sulfur ecosystem remains a key goal for the field.

Observations of CS in astrophysical environments are typically made through two main methods: {radio (millimeter)} rotational transitions in the ground X$^1\Sigma$ state and near-ultraviolet (NUV) fluorescence in the A$^1\Pi$-X$^1\Sigma$ system. These techniques have been applied to comets within our solar system to investigate potential precursor molecules, such as CS$_{2}$, OCS, and H$_{2}$CS{, all of which have been detected in comae \citep{woodney1997detection,woodney1997sulfur,dellorusso1998,jackson2004,biver2015,leroy2015inventory,calmonte2016sulfur,saki2020}}. However, a factor of 2-5 discrepancy arises between the CS/H$_2$O abundances determined from NUV and radio observations of {a comet during the same apparition, with NUV observations consistently producing higher CS production rates. This holds true for almost all comets with a CS detection from HST/STIS and the radio thus far: 46P/Wirtanen in the 2019 apparition \citep{noonan2023evolution,biver2023coma}, 67P/Churyumov-Gerasimenko in the 2021 apparition \citep{noonan2023evolution,biver2023coma}, C/2009 P1 (Garradd) in 2012 \citep{biver2012garradd,yang2012garradd,gicquel2015,feldman2018}, and C/2014 Q2 (Lovejoy) in early 2015 \citep{biver2016,feldman2018}. This is notably not the case for measurements of C/1995 O1 (Hale-Bopp) pre-perihelion, which was more active, observed at larger heliocentric distances, and also observed with the narrow aperture IUE LWR, and HST FOS instruments, not STIS \citep{biver2002,weaver1997}. Intriguingly, in the post-perihelion period, when STIS was used for NUV measurements rather than IUE or FOS, began to show a higher derived CS parent production rate \citep{weaver1999}. \edit{Only a portion of this discrepancy can be attributed to a lower water production rate Q(H$_2$O) derived from OH emissions in the NUV used for the CS/H$_2$O abundances; simultaneously obtained NUV and radio measurements typically result in a derived NUV Q(H$_2$O) that is a factor of 1 - 2 lower than the radio \citep{weaver1999}.} Moreover, the CS distribution observed for many comets appears spatially extended in the radio, in sharp contrast to the steep radial profiles observed at UV wavelengths.} These variations complicate direct comparisons across the wavelength regions and require an evaluation of the excitation efficiencies employed in calculating both the total molecular column densities and production rates.

CS was first detected in comet C/1975~V1 (West) in 1975 \citep{Smith1980Comet} {at UV wavelengths,} and to derive its general abundance, the authors calculated the fluorescence efficiency (or $g$-factor) of the A$^1\Pi$-X$^1\Sigma$ $(0,0)$ band, presumably with only one vibrational level in each electronic state, at a heliocentric/geocentric distance of 0.52~au. The following years saw investigations of the band structure {in the spectrum of C/1979 Y1 (Bradfield) to derive an X$^1\Sigma~(v=0)$}  Boltzmann temperature of 70 K~\citep{jackson1982production}, and found that the ratios of the $(0,0)$, $(1,0)$, and $(0,1)$  bands of the A$^1\Pi$-X$^1\Sigma$ system are well described by resonant fluorescence of solar radiation. The spatial {distribution} of S and CS emissions observed in comet C/1979 Y1 (Bradfield), as discussed in \citet{jackson1982production}, suggested CS$_{2}$ as a parent molecule of CS due to its short lifetime and corresponding small scale length{, a suggestion that was ultimately bolstered by the detection of the molecule in the NUV and visible spectrum of comet 122P/de Vico~\citep{jackson2004}}. Following this, \cite{Sanzovo1993CS} recalculated the $g$-factor for the {A$^1\Pi$-X$^1\Sigma$ $(0,0)$} band and determined production rates for 15 comets. Subsequent papers referred to a fluorescence efficiency provided through a private communication with few details on how it was derived (e.g.,~\citealt{Noll1995,Stern1998,noonan2023evolution}). These derived $g$-factors are summarized in Table \ref{tab:previous_CS_gf}. {Even for models that used the same solar reference spectrum~(e.g. \cite{Sanzovo1993CS} and \cite{Noll1995}), there is a variation of order a factor of 2 after correcting for heliocentric distance} between the derived fluorescence efficiencies, and the effect of the comet's orbital velocity with respect to the Sun on the fluorescence efficiency (Swing's effect) has not been evaluated. As noted in \cite{noonan2023evolution}, improved fluorescence efficiencies for the NUV bands of CS may contribute to a resolution of the NUV/radio CS abundance discrepancies. To enhance the accuracy of CS column density measurements derived from NUV observations, it is necessary to rigorously evaluate the fluorescence efficiencies of the A$^1\Pi$-X$^1\Sigma$ bands.

\begin{table*}[]
    \centering
    \begin{tabular}{l|c|l|c}
        Publication & Heliocentric & Solar Spectral &  CS A$^1\Pi$-X$^1\Sigma$ \\
        & Distance (au)  & Reference & $(0,0)$ $g$-factor \\
        & & & (photons$^{-1}$ s$^{-1}$ molecule$^{-1}$)\\
        \hline
        \cite{Smith1980Comet} & 0.52 & \cite{Malitson1960solar} & 3.8~$\times$ 10$^{-3}$\\ 
        \cite{jackson1982production} & 1.0 & \cite{Kohl1978solar}& 7~$\times$ 10$^{-4}$\\
        \cite{Sanzovo1993CS} & 0.52 & \cite{Kohl1978solar, 1983umd..rept...44A}& 3.9~$\times$ 10$^{-3}$\\
        \cite{Noll1995} & 1.0 & \cite{1983umd..rept...44A} &5.8~$\times$ 10$^{-4}$\\
        \cite{Stern1998} & 1.0 & \cite{1983umd..rept...44A} & 5~$\times$ 10$^{-4}$\\\hline
        
    \end{tabular} 
    \caption{CS A$^1\Pi$-X$^1\Sigma$ $g$-factors from 1980 - Present}
    \label{tab:previous_CS_gf}
\end{table*}

In this paper we perform new computational modeling of the fluorescing CS A$^1\Pi$-X$^1\Sigma$ system using the FlorPy code \citep{Bromley2023}, utilizing an updated set of transition rates benchmarked against an electron impact spectrum produced from e+CS$_{2}$ collisions. We review our modeling methodology, benchmarking experiment, and discuss for the first time the Swings effect for the CS A$^1\Pi$-X$^1\Sigma$ system. We conclude by reviewing astrophysical observations that can benefit from the improved model and review open questions as to the parentage of the CS radical.

\section{Fluorescence Modeling}
\subsection{Theoretical Approach}
Once an atom or molecule is produced in a cometary coma, the emitter begins to absorb and re-emit sunlight. Assuming a species is formed in the lowest energy (ground) state, absorptions of solar photons occur and re-distribute population amongst its energy levels. The level populations $n_i$ can be computed from a series of coupled equations written as
\begin{equation}\label{eq:detailed_balance}
\frac{dn_i}{dt} = -n_i\sum^N_{i\neq{k}}(A_{i\rightarrow{k}} + {\rho}B_{i\rightarrow{k}}) + \sum^N_{j\neq{i}}n_j(A_{j\rightarrow{i}} + {\rho}B_{j\rightarrow{i}})
\end{equation}
where $A$ is the Einstein $A$ coefficient (the `transition rate'), $B$ are the Einstein coefficients for absorption and stimulated emission, and $\rho$ is the radiation field. {We include $N = 909$ rovibronic levels in our model, interconnected by 16019 transitions, with $n_i$ denoting each level described by a series of electronic, vibrational, and rotational quantum numbers and labels. The system of equations is completed by the replacement of one equation with the normalization condition $\sum_i n_i = 1$~\citep{Bromley2021,Magnani1985,Bromley2023}. After sufficiently large times, the level populations reach a steady-state, fluorescence equilibrium ($dn/dt = 0$), whose solution is tractable by numerous computational methods~(see discussion in \citealt{Bromley2021}).} From the level populations, the fluorescence efficiencies, or $g$-factors, are computed (in units of Joules~s$^{-1}$~molecule$^{-1}$) as
\begin{equation}\label{eq:g-factor}
g_{ji} = \frac{hc}{\lambda_{ji}}n_jA_{j\rightarrow{i}}
\end{equation}
where $hc/\lambda_{ji}$ is the photon energy of the transition, $n_j$ is the equilibrium population of the upper level involved in the transition, and $A_{j\rightarrow{i}}$ is the transition rate. $g$-factors are also commonly reported in units of photons s$^{-1}$ molecule$^{-1}$, computed with the omission of $hc/\lambda$ in Eq.~\ref{eq:g-factor}. 

The fluorescence rate equations (Eq.~\ref{eq:detailed_balance}) describe the level populations, assuming an environment dominated by  photoabsorption and radiative decays. {Depending on the distance at which the species is produced, some of the energy levels of a species may thermalize to some degree with the neutral coma.} For example, \cite{jackson1982production} found that the distribution of rotational states in CS, as inferred from high-resolution spectra of the A$^1\Pi$-X$^1\Sigma$ $(0,0)$ band, implied some degree of thermalization {in X$^1\Sigma$ that may explain the observed band shape. This complication may impact the magnitude of the band luminosities, defined as the sum of the $g$-factors of all transitions within a given band, and contribute to the ongoing discrepancy between NUV and and/or radio-derived CS abundances.}

{Theoretically,} the thermalization of the X$^1\Sigma(v=0)$ ground state can be readily incorporated into Eq.~\ref{eq:detailed_balance}. For a given temperature $T$, the (relative) level populations within X$^1\Sigma$ can be computed as a Boltzmann distribution characterized by $T$. Eq.~\ref{eq:detailed_balance} thus reduces to $N - m$ linearly coupled equations, where $N$ is the total number of levels, and $m$ is the number of known (constrained) levels described by $T$. {The system of equations can be solved using the same methods as the pure fluorescence approach with the known $n_j{\rho}B$ terms now computed from the constrained (fixed) level populations $n_j$. A similar approach was used by \cite{Kim1990} to compute time-dependent emission from S$_2$. In the present manuscript, we solve both the pure fluorescence} (Eq.~\ref{eq:detailed_balance}) and partially-thermalized cases to investigate whether such effects dramatically alter the fluorescence properties of CS.

For the purposes of computing fluorescence efficiencies of CS, we utilize the Python-based fluorescence code `FlorPy' described in \cite{Bromley2021,Bromley2023}. In practice, the rate equations are agnostic to the details of the atomic or molecular structure, and rely solely on transitions connecting levels described by their energies and total angular momenta $J$. Transitions themselves are described by a wavelength (or equivalently, an energy) with an associated transition rate, {i.e. an Einstein $A$ coefficient}, in units of s$^{-1}$. Einstein B coefficients are computed from the Einstein $A$ coefficients. We utilize here the solar spectrum described in \cite{Bromley2023}, where the fluxes pumping the A$^1\Pi$-X$^1\Sigma$ transitions of CS are from the re-calibration efforts described in \cite{Coddington2021}, with the original high-resolution spectrum deriving from \cite{Anderson1991}. Pumping of the infrared transitions of CS is computed from high-resolution infrared spectra generated with the Planetary Spectrum Generator~(PSG;~\citealt{PSG}). The infrared solar spectrum utilized in the PSG derives from a series of high-resolution measurements conducted by the {Atmospheric Chemistry Experiment (ACE)} mission~\citep{Hase2021}.

In \cite{Bromley2023} we presented a pipeline that was developed to produce fluorescence emission models of diatomic molecules, with particular application to CO$^+$. This pipeline relied on existing molecular constants and employed the PGOPHER program~\citep{pgopher} to model the vibronic structure of the cation. We apply an analagous approach here to construct a line list of CS using molecular constants reported in \cite{Bergeman1981} and the PGOPHER program~\citep{pgopher} for the X$^1\Sigma$ ground state and the first excited state A$^1\Pi$. {For the A$^1\Pi(v=0)$ levels, we use the updated molecular constants from \cite{Kuroko2022}.} Within each electronic state, vibrational levels up to $v_\textrm{max} = 8$ and rotational levels up to $N_\textrm{max} = 100$ are retained. Within each vibrational level, we utilize the experimentally-derived constants for molecular rotation $(B)$ and the centrifugal distortion constants $D$. {As noted by \cite{Bergeman1981}, the A$^1\Pi$-X$^1\Sigma$ bands of CS are subject to substantial perturbations from interactions with triplet states. The impact of these perturbations on both the spectra and spatial profiles of CS are discussed in more detail in Section~\ref{sec:discussion}.} 

%In most reported laboratory and astronomical observations of CS in the ultraviolet, the spectral resolution does not permit clear separation between rotational lines, owing to the small rotational constants~($B \sim 0.8$~cm$^{-1}$). 

Previous calculations on diatomics relevant to cometary science demonstrated that including both rovibronic transitions between electronic states and rovibrational transitions within electronic states are required to accurately model the level populations~\citep{Magnani1985,Schleicher2010,Bromley2023}. In the present calculations, rovibronic transitions between all included vibrational levels of the X$^1\Sigma$ and A$^1\Pi$ electronic states are retained, as well as vibrational and rotational transitions within the X$^1\Sigma$ ground state.  Einstein $A$ coefficients for the transitions of CS are sourced from the calculations of \cite{Xing2020}, and benchmarked against experimental measurements as described below. Energy level and transition line lists, which are used as inputs to the fluorescence model, are created from parsing the outputs of the PGOPHER calculations, which are built on the constants described previously and the newly-benchmarked transition rates, described further below.

\subsection{Experimental Benchmarking}
In the rate equations of fluorescence equilibrium, fluorescence efficiencies and their resulting band luminosities (produced by sums over each band's transitions) depend on the transition rates. Further, fluorescence efficiencies and band luminosities are proportional to the Einstein $A$ coefficient (Eq.~\ref{eq:g-factor}). \cite{Xing2020} noted significant discrepancies between computed lifetimes of CS vibrational levels and those reported from  experimental efforts. To improve the accuracy of the computed data utilized in our models we re-normalized the theoretical lifetimes to the experimentally-measured lifetimes reported by \cite{Mahon1997}, who used laser-induced fluorescence to selectively excite bands of CS and measure the resulting radiative lifetimes. For $v > 4$, for which no measured radiative lifetimes are available, a linear fit ($R^2 = 0.997$) to the $v < 4$ lifetimes of \cite{Mahon1997} is used to extrapolate out to higher $v$. The linear fit is used to predict experimental lifetimes for $v > 4$, which are $\sim$30\% larger than those originally reported in \cite{Xing2020}.

\begin{figure*}[t]
    \centering
   \includegraphics[scale=0.5]{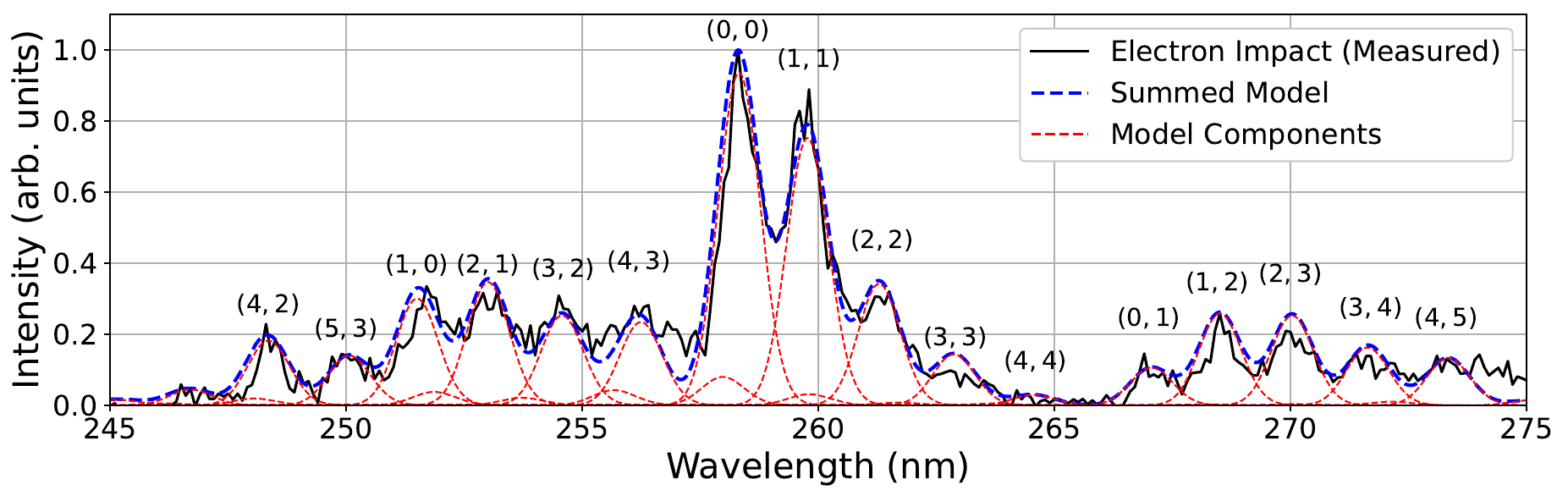}
    \caption{{Comparison of the theoretical model (blue; summed over model components (red)) and electron impact spectrum produced by 100 eV electrons incident on CS$_2$ gas (black). Emission bands of the CS A$^1\Pi$-X$^1\Sigma$ system are labeled by $(v',v'')$, where $v'$ and $v''$ are the upper and lower vibrational numbers, respectively. The modeled spectrum is generated from benchmarked transition rates and vibrational populations derived from the strongest feature associated with each A$^1\Pi$ vibrational level. The electron impact spectrum is corrected for the sensitivity of the optical elements, and a minor wavelength shift is applied to the theoretical spectrum to account for wavelength calibration errors.}}\label{fig:labcomp}
\end{figure*}

For a given vibrational level labeled by $v'$, the branching ratios to lower vibrational levels, labeled by $v''$, are described by Franck-Condon factors. While these values are reported in various theoretical efforts, we opted to benchmark the relative transition rates (Franck-Condon factors) extracted from the tables of \cite{Xing2020} against experimental measurements. For this purpose, the emission spectrum of CS was recorded from crossed-beam electron impact experiments on CS$_2$ gas at the Laboratory of Electron Induced Fluorescence at Comenius University in Bratislava, Slovak Republic in Summer 2023. Details of the experimental apparatus are publicly available~(see \cite{Danko_2013,Bodewits_2019,Bromley2023}, and references therein). In short, an energy-monochromatized electron beam is crossed perpendicularly with a neutral beam of CS$_2$ gas inside a vacuum chamber (base pressure $\sim 10^{-8}$ mbar). Experiments were conducted in the single-collision regime with chamber pressures during gas injection not exceeding $10^{-4}$ mbar. Photons from decaying collision products (CS, CS$_2$, and CS$_2^+$) were recorded by a Czerny-Turner monochromator equipped with a UV-sensitive photomultiplier tube. The optical response of the apparatus was established by previous measurements of a known system, H$_2$ $a^3\Sigma_g^+$-$b^3\Sigma_u^+$, and the thermal blackbody produced by a heated tungsten filament. An example of the resulting laboratory spectrum is shown in Figure~\ref{fig:labcomp} for a collision energy of 100~eV.

At electron energies above the threshold of CS (A$^1\Pi$) formation from CS$_2$ ($\sim$9.27~eV), a large number of CS A$^1\Pi$-X$^1\Sigma$ bands are excited between 250 -- 280~nm. Our observed relative intensities of the many CS bands agree well with those reported in \cite{Ajello1971}. In order to benchmark the vibrational transition rates reported in \cite{Xing2020}, an iterative procedure  was applied (described in the Appendix of \citealt{Bromley2023}). First, the relative intensities of all bands were extracted by either an integral over the spectral extent of each band, or in the case of strong blends (the case for most bands of CS), was manually adjusted to match the laboratory spectrum. To refine the handling of the many blended features, vibrational level populations are derived from the strongest bands of each A$^1\Pi$ vibrational level, and manually adjusted to match the observed spectrum. The relative transition rates are then normalized to the lifetimes described above. We note that the rotational structure of the bands is not resolved, though the lifetimes appear weakly sensitive to the rotational quanta $J$~\citep{Xing2020}. Further, for the purposes of the present benchmarking, a rotational temperature of $\sim$300~K was assumed. It is possible that the rotational distribution is non-thermal, but a more rigorous derivation of the rotational populations cannot be derived at the present spectral resolution. Lastly, the level populations are re-derived using the updated transition rates, and the process was repeated until sufficient agreement with the laboratory spectrum was achieved. In total, this process leads to changes of order 20 -- 30\% in transition rates of the strongest bands, and up to factors of $\sim$2 in weak transitions. Figure~\ref{fig:labcomp} shows a comparison of the laboratory spectrum at 100~eV electron energy and our final, benchmarked model built on the re-calibrated transition rates. The agreement with the experimental spectrum is excellent and lends confidence to our use of the benchmarked transition rates in computing fluorescence efficiencies.

\section{Results}
\subsection{Band Luminosities}

\begin{table}
    \centering
    \begin{tabular}{|c|c|c|c|}
    \hline
  {$v'$}~$\backslash$~{$v''$}&0&1&2\\ \hline
  &257.73&266.47&275.72\\
0&3.93e-22&4.3e-23&1.21e-24\\
&5.06e-04&5.79e-05&1.68e-06\\\hline
&250.94&259.21&267.96\\
1&1.76e-23&4.28e-23&1.29e-23\\
&2.23e-05&5.59e-05&1.74e-05\\\hline
&244.58&252.44&260.72\\
2&8.39e-26&1.75e-24&1.68e-24\\
&1.03e-07&2.22e-06&2.20e-06 \\ \hline
    \end{tabular}
    \caption{Fluorescence Efficiencies of the CS (A$^1\Pi$-X$^1\Sigma$) bands at $r_\textrm{h} = 1$~au and $v_\textrm{h} = 0$~km~s$^{-1}$. Fluorescence efficiencies are provided as a function of upper ($v'$; rows) and lower ($v''$; columns) vibrational quantum numbers, respectively. For each band, the following values are provided (top to bottom): vacuum wavelength (nm) for the R(0) line within the band, the band luminosities (summed over all transitions) in units of J~s$^{-1}$~molecule$^{-1}$, and band luminosities in units of photons~s$^{-1}$~molecule$^{-1}$. Band luminosities for all bands (including $v',v'' > 2$) are provided as supplementary files.}\label{tab:bandlums_a-x} 
\end{table}

Previous fluorescence efficiencies of the A$^1\Pi$-X$^1\Sigma$ (0,0) band of CS range from $1\times10^{-4}$ to $5.8\times10^{-4}$ photons s$^{-1}$ molecule$^{-1}$ at a heliocentric distance of 1 au~\citep{Smith1980Comet,jackson1982production,Noll1995,Stern1998} (see Table \ref{tab:previous_CS_gf}). As a first test of our model, we computed fluorescence efficiencies of the CS ultraviolet bands at a heliocentric distance of $r_\textrm{h}$ = 1~au and a heliocentric velocity of $v_\textrm{h}$ = 0 km~s$^{-1}$. The resulting band luminosities, summed over the transitions in each band, are provided in Table~\ref{tab:bandlums_a-x}. The luminosity of the $(0,0)$ band, $5.07\times10^{-4}$ photons s$^{-1}$ molecule$^{-1}$, is in good agreement with the value reported in \cite{Noll1995}. Both \cite{Noll1995} and the present work utilize the molecular constants reported in \cite{Bergeman1981}. However, the extent of the rovibrational structures utilized in \cite{Noll1995} is not clear. We find that bands from $v' > 2$ (not shown) have negligible band luminosities and are omitted from Table~\ref{tab:bandlums_a-x}. The present values affirm that neither the updated solar spectrum, nor the large rovibrational level structure utilized here, have introduced large changes to the baseline g-factors.

As most UV comet observations have utilized low spectral resolution ($R\sim1000$), the topic of blends must be addressed. In the laboratory, the strong $(0,0)$ band is blended with the neighboring $(1,1)$ band. In fluorescence, the equilibrium populations of the X$^1\Sigma (v=1)$ rotational levels are small compared to $(v=0)$. Consequently, the fluorescence band luminosity of the $(1,1)$ band is less than 1/10\textsuperscript{th} that of the $(0,0)$ band, and the overlap between the bands is negligible. It can be concluded that any observed flux attributed to the CS $(0,0)$ band in comet observations is  free of blends with other bands of the A$^1\Pi$-X$^1\Sigma$ system.

Lastly, we consider the thermalization of the X$^1\Sigma$ ground state. As CS is produced in the inner coma, it is plausible that {the ground rovibronic levels are thermalized to some extent}, and may be described by a Boltzmann distribution~\citep{jackson1982production}. The NUV spectrum produced from pumping of a thermalized ground state population is likely to excite a wider range of {NUV} transitions than pure fluorescence alone. We pursued calculations of band luminosities assuming X$^1\Sigma(v=0)$ thermalization temperatures of 30, 70, 100, and 300 Kelvin. The resulting luminosities of the $(0,0)$ band differ by at most 4\%, though the band shape can change dramatically with ground state temperature. Thus, for the purposes of deriving CS abundances, pure-fluorescence band luminosities are of sufficient accuracy.

\subsection{Cometary Spectra Benchmarks}

\begin{figure}
    \centering
   \includegraphics[scale=0.8]{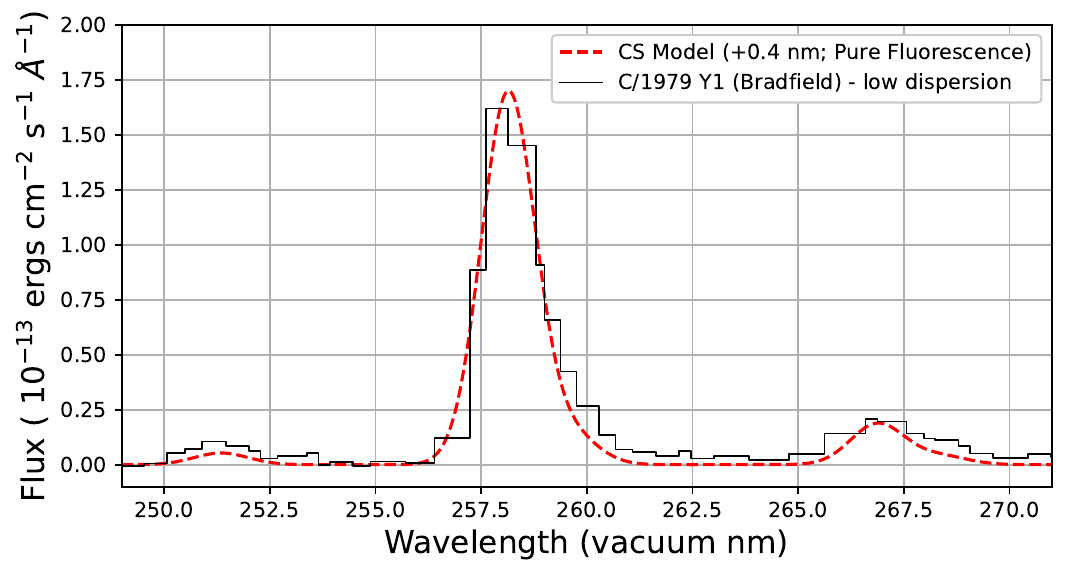}
    \caption{Present CS fluorescence model versus low-dispersion spectra of  C/1979 Y1 (Bradfield) (observation ID LWR 6008), collected on 10 Jan. 1980 at $r_\textrm{h}$ = 0.711 au  from~\citep{Ahearn1980}. The A$^1\Pi$-X$^1\Sigma$ bands of CS (left to right) $(1,0)$, $(0,0)$, and $(0,1)$ show good relative agreement with the fluorescence model.}\label{fig:bradfield_lowres}
\end{figure}

As a benchmark test for these models, we apply them to both high and low dispersion spectra of  C/1979 Y1 (Bradfield), collected with the International Ultraviolet Explorer (IUE). Figure~\ref{fig:bradfield_lowres} shows a comparison of the low-resolution IUE spectrum (LWR 6008, 1980 Jan. 10) of \cite{Ahearn1980} at $r_\textrm{h} = 0.711$~AU, and our pure-fluorescence model. Good agreement is observed between the cometary CS and the modeled spectra of the A$^1\Pi$-X$^1\Sigma$ $(0,0)$, $(0,1)$, and $(1,0)$ bands. {We also explored equilibrium models assuming a thermalized ground state at varying temperatures, and observed negligible changes in the theoretical spectrum at low spectral resolution. Additionally, a wavelength shift of +0.4~nm is applied to the theoretical spectrum. This shift is approximately 1 pixel width, and is assumed to correct for inaccuracies in the wavelength calibration. The wavelength scale, i.e. in vacuum or converted to in-air wavelengths, is also not clear from \cite{Ahearn1980}. We assume here that these data are presented in vacuum.}

\begin{figure}
    \centering
   \includegraphics[scale=0.8]{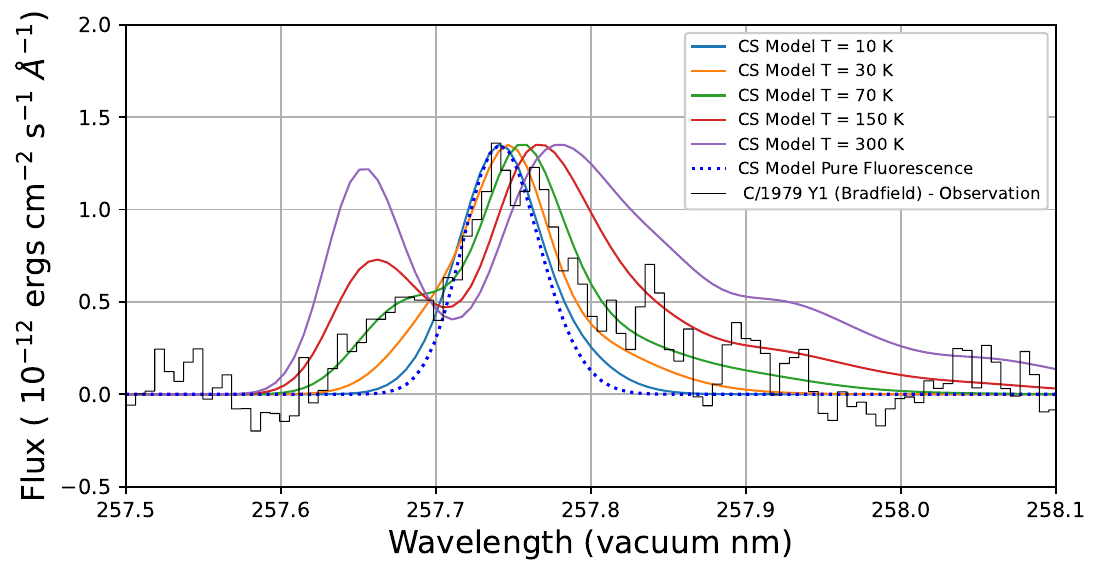}
    \caption{Present CS fluorescence models versus high-dispersion spectra of  C/1979 Y1 (Bradfield) (observation ID LWR 6757), collected on 24 Jan. 1980 at $r_\textrm{h}$ = 0.925 au~\citep{jackson1982production}. The shape of the $(0,0)$ band is well described by the fluorescence model with fixed X$^1\Sigma(v=0)$ populations described by $T = 70$~K~\citep{jackson1982production}.}\label{fig:bradfield_highres}
\end{figure}

{High-dispersion observations of  C/1979 Y1 (Bradfield), described in \cite{jackson1982production} were collected on 24 Jan. 1980. Figure~\ref{fig:bradfield_highres} provides a comparison of the IUE spectra and our spectral modeling, with both pure-fluorescence and thermalized ground state models between 10 and 300~K shown. The spectra were processed using the standard IUE pipeline, and are shown with a vacuum wavelength scale, corrected for all velocity shifts between the sun, comet, and observer (IUE). Pure-fluorescence models are unable to account for the enhanced emission in the red and blue wings of the band. With increasing temperature, greater agreement is found in the red wing of the band, with poorer agreement in the blue wing. Even in the model with the greatest agreement ($T = 70$~K), an apparent deficiency in flux, red-ward of maximum intensity, remains apparent.  A similar approach was utilized by \cite{jackson1982production}, and found similar agreement with the observed spectrum.}

{The wavelength position of peak flux in the $(0,0)$ band shifts toward the red with increasing temperature. At higher thermalization temperatures, a (negative) wavelength shift is required to match the observed positions. Previous modeling efforts by \cite{Swamy1993} also observed similar shifts in the wavelength of peak flux in time-dependent models. They utilized time- and density-dependent models, assuming a constant outflow velocity, and solved for the level populations and $g$-factors from initially `hot' distributions of energy levels.  Integrated over the relevant fields of view, they achieved good agreement with the NUV spectra of C/1979 Y1 (Bradfield), 21P/Giacobini-Zinner, and to a lesser extent, 1P/Halley. We note that \cite{Swamy1993} primarily addressed the CS A$^1\Pi$-X$^1\Sigma$ $(0,0)$ band shape, and does not report the effect on the absolute flux of the band, i.e., the band luminosity that is historically used to derive CS abundances.}

{The present models, both pure-fluorescence and those assuming only a thermalized ground state, find agreement with low- and high-resolution spectra of C/1979 Y1 (Bradfield), respectively. Importantly, the use of pure-fluorescence versus thermalized ground state models has a negligible impact on the band luminosities, and consequently, a negligible impact on the derivation of CS abundances with the present models.} 

{In the following, we expand the scope of our models beyond comparisons to single objects, and investigate whether heliocentric distance, velocity, and time-dependent effects lead to changes in band luminosities which may explain the apparent enhancement of NUV-derived CS abundances.}

\subsection{Heliocentric Distance, Velocity, and Time-Dependent Effects}
Once produced, CS has a comparable photochemical lifetime to H$_2$O, of order $10^{5}$~seconds against {photodissociation} at 1~au~\citep{Meier1997,Heays2017}. {The use of equilibrium fluorescence efficiencies in deriving accurate abundances from observed fluxes necessitates the validity of fluorescence equilibrium, i.e. that CS produced in the inner coma achieves equilibrium well within the field of view.} We checked the validity of assuming fluorescence equilibrium by computing a time-dependent solution of the fluorescence rate equations~(Eq.~\ref{eq:detailed_balance}). 

We utilize the time-dependent matrix method described previously in \cite{Bromley2023}, assuming first that the CS is initially formed in the lowest energy state. From time-dependent calculations of the populations and $g$-factors, we find that the luminosity of the $(0,0)$ band evolves from a value $\sim10\%$ greater than its equilibrium value at $t = 0$, and asymptotically approaches within $5\%$ of the equilibrium fluorescence efficiency within $t = 10^{3} - 10^{4}$ seconds at 1~au. This suggests that any {NUV} emission from CS initially produced in the ground rovibrational state is practically indistinguishable from a population of CS in fluorescence equilibrium. {It can be concluded that the band luminosity of the A$^1\Pi$-X$^1\Sigma$ $(0,0)$ band in CS freshly produced in the ground rovibronic state is negligibly impacted by the assumption of fluorescence equilibrium.}

CS$_2$, the often-attributed parent of CS~\citep{noonan2023evolution} produces CS in highly excited vibrational and rotational states upon photodissociation~\citep{McCracy1985}. \cite{Swamy1993} found that such distributions relax toward fluorescence equilibrium much slower than CS initially formed in the ground rovibrational state. We explored similar possibilities in the present time-dependent models, assuming initial conditions described by Boltzmann distributions with a range of temperatures from 50 to 1000~K. 

\begin{figure*}
    \centering
   \includegraphics[scale=0.5]{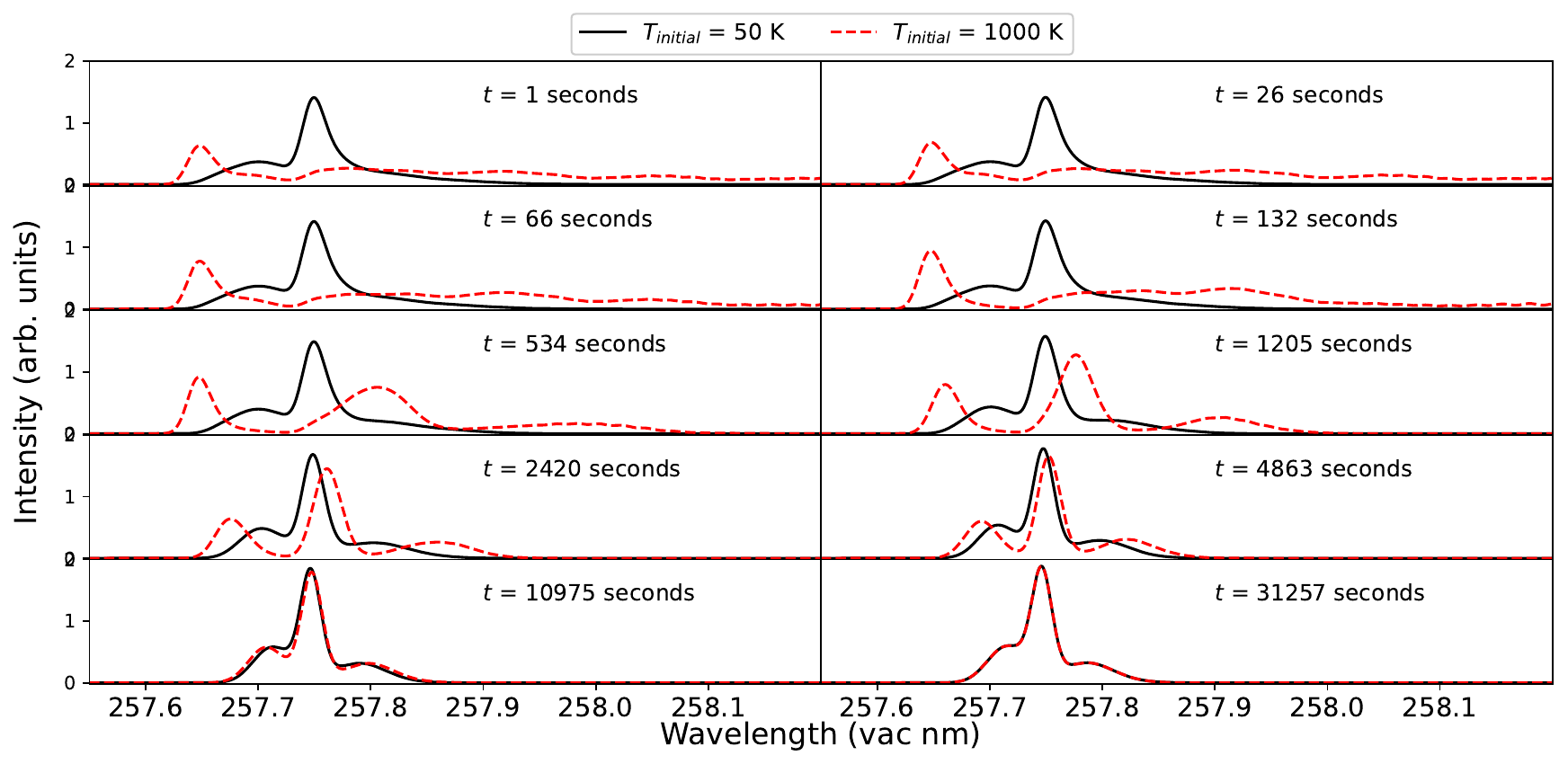}
    \caption{Computed spectrum of CS A$^1\Pi$-X$^1\Sigma~(0,0)$ from two time-dependent pure-fluorescence models with different initial conditions. Orbital conditions of $r_\textrm{h} = 1$~au and $v_\textrm{h} = 0$~km~s$^{-1}$ are assumed. The two models utilize different initial conditions, where the levels at $t = 0$ are given by Boltzmann distributions characterized by $T = 50$~K (black curve) and $T = 1000$~K (red curve). Transition-resolved $g$-factors are convolved with Gaussians with 0.02~nm FWHM to produce the presented profiles.}\label{fig:td_bandshapes}
\end{figure*}

Figure~\ref{fig:td_bandshapes} shows the time-evolution of the two extremes, where each population of CS is formed with X$^1\Sigma~(v=0)$ levels described by a Boltzmann temperature of 50~K and 1000~K. The population of CS formed with $T = 50$~K relaxes to fluorescence equilibrium rapidly. The population of CS formed at $T = 1000$~K, for which the most-populated X$^1\Sigma(v=0)$ {rotational level} peaks around $J=20$, evolves to the same equilibrium over longer timescales of order $10^4$~seconds. The distribution of CS formed initially at 1000~K has apparent separation between the R branch (blueward of 257.7~nm) and the Q ($\sim$257.8~nm) and P (redward of 257.8~nm) branches. With increasing time, the two branches merge, with a noticeable peak increasing in intensity and shifting from the wavelength position of the P branch towards that seen in fluorescence equilibrium. {Despite the marked difference in band shape between these two models as a function of time, the band luminosities in both models agree within several percent. While time-dependent effects may be necessary to understand the band shape, they do not contribute meaningfully to the NUV/radio abundance discrepancy.}

%{If such a distribution were initially formed and evolving in the inner coma, one would expect a shift in the wavelength position in the peak flux of the $(0,0)$ band when integrated over the coma, as was observed in comparisons to IUE observations of  C/1979 Y1 (Bradfield)~(Figs.~\ref{fig:bradfield_lowres}-\ref{fig:bradfield_highres}).}

{Summarizing, we observe shifts in the wavelength position of the peak flux of the $(0,0)$ band using the present time-dependent models. As we have neglected collisions and still recover the behaviors identified by the density- and time-dependent models of \cite{Swamy1993}, we conclude that the evolution of the band shape is strongly determined by the time evolution of the distribution of initial states, rather than collisional effects. Additional modeling work may be explored to incorporate collisional effects in an attempt to further constrain the distribution of initially-populated states upon dissociation to CS. Despite the missing physical effects in our fluorescence equilibrium models, i.e. density- and time-evolution effects of initially `hot' states, the computed band luminosities are similar to those generated from either partially-thermalized models, or fully time-dependent models with different initial conditions. In total, the present results suggest that existing literature which utilized fluorescence equilibrium $g$-factors to derive CS abundances from observed fluxes are secure against improvements in computed fluorescence efficiencies. Other factors must contribute to the 2 - 5$\times$ increase in NUV-derived abundances compared to radio observations.}

As noted in the case of CO$^+$~\citep{Bromley2023}, {the band shapes in fluorescence equilibrium are determined by the} competition of absorption rates between electronic states and radiative rovibrational transition rates within the ground electronic state. In the case of CS, the infrared transitions are relatively weak, with the strongest IR band, the X$^1\Sigma$-X$^1\Sigma$ ($1,0)$ band between 7.8 - 7.9~$\mu$m, having a band luminositiy $\sim{1}/100^{\textrm{th}}$ that of the A$^1\Pi$-X$^1\Sigma$ $(0,0)$ band. {Given the relative weakness of the infrared emission, a minor change in the band emission beyond a typical $1/r_h^2$ dependence may be expected. This assertion is discussed later.}

\begin{figure}
    \centering
   \includegraphics[scale=0.5]{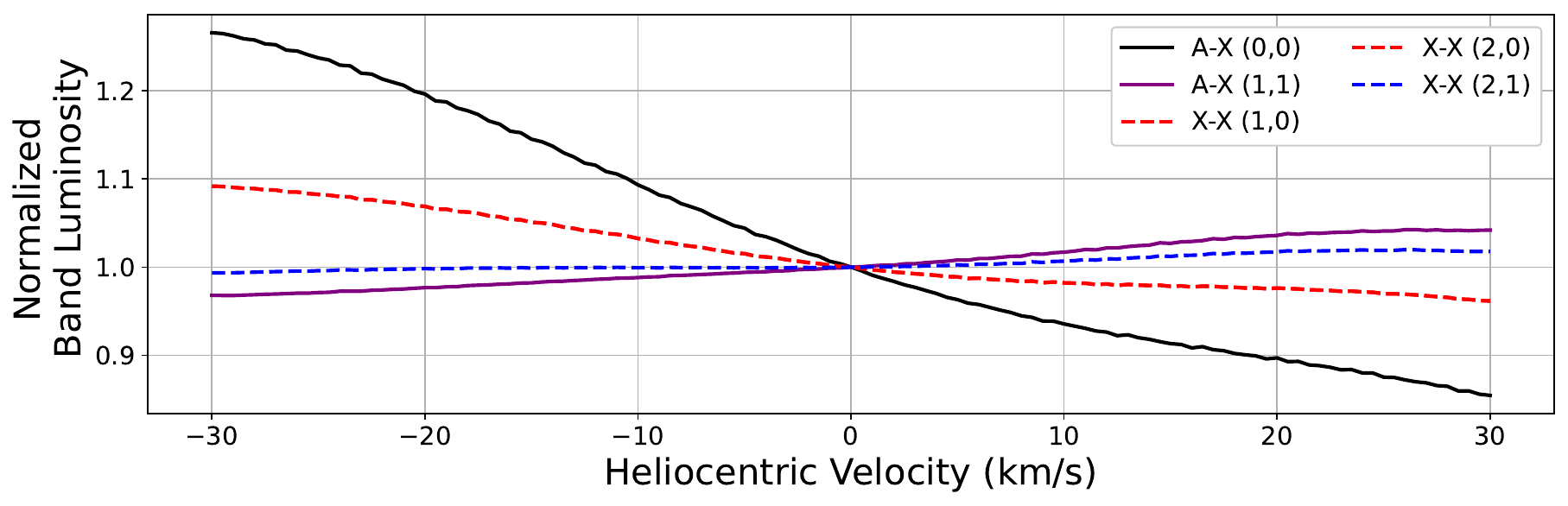}
    \caption{Heliocentric velocity dependencies of {four representative bands of CS.} Each band is normalized to a value of 1 at $v_\textrm{h} = 0$~km~s$^{-1}$.}\label{fig:veldep}
\end{figure}

Large variations in fluorescence efficiency— exceeding factors of two—with respect to heliocentric velocity have been observed in diatomics such as OH \citep{Schleicher1988} and CN \citep{Schleicher2010}. However, reports on how CS band luminosities vary with heliocentric velocity are currently lacking and we therefore explored the potential for similar trends in CS. Figure~\ref{fig:veldep} shows the band luminosities of the two strongest A$^1\Pi$-X$^1\Sigma$ bands of CS as a function of heliocentric velocity. The band luminosity of the strong $(0,0)$ band is relatively insensitive to heliocentric velocities for $|v_\textrm{h}| < 20$~km~s$^{-1}$. At large heliocentric velocities $|v_\textrm{h}| > 20$~km~s$^{-1}$, variations in excess of 10\% are apparent. The neighboring weak $(1,1)$ band appears insensitive to velocity, changing by at most 3\% over the present velocity space of {-30 to +30} km~s$^{-1}$, a reasonable parameter space for solar system comets.

{Similarly, previous studies have commonly computed the fluorescence efficiencies of CS at $r_\textrm{h} = 1$ au and scaled the fluorescence efficiencies to the necessary distance(s) by a factor of $1/r_\textrm{h}^2$. Band luminosities, {i.e. the total emission of a particular band, may be insensitive to heliocentric distance, even while the distribution of features within a band can change dramatically, as was observed in e.g. CO$^+$~\citep{Bromley2023}.} In some cases, such as CN~\citep{Schleicher2010}, the band luminosities are sensitive to both heliocentric distance and velocity, and a simple rescaling by $1/r_h^2$ may not be valid.} As discussed previously, the A$^1\Pi$-X$^1\Sigma$ bands of CS are weakly sensitive to heliocentric velocity. To quantify the extent to which scaling $g$-factors from 1 au to $r_\textrm{h}$ is valid, we explicitly calculated the fluorescence efficiencies up to $r_\textrm{h} = 10$~au {by including the distance scaling of the radiation field in Eq.~\ref{eq:detailed_balance}.} Within the range of typical heliocentric distances ($<$ few au) where UV observations have captured CS ultraviolet emission, the band luminosities changed by at most 2\% between calculations re-scaled from 1~au, and those that explicitly account for the distance dependence of the absorption rates in the rate equations~(Eq.~\ref{eq:detailed_balance}). {In total, neither heliocentric velocity or distance effects can explain the relative enhancement of NUV-derived CS abundances when compared to radio observations.}

\begin{figure*}
    \centering
   \includegraphics[scale=0.55]{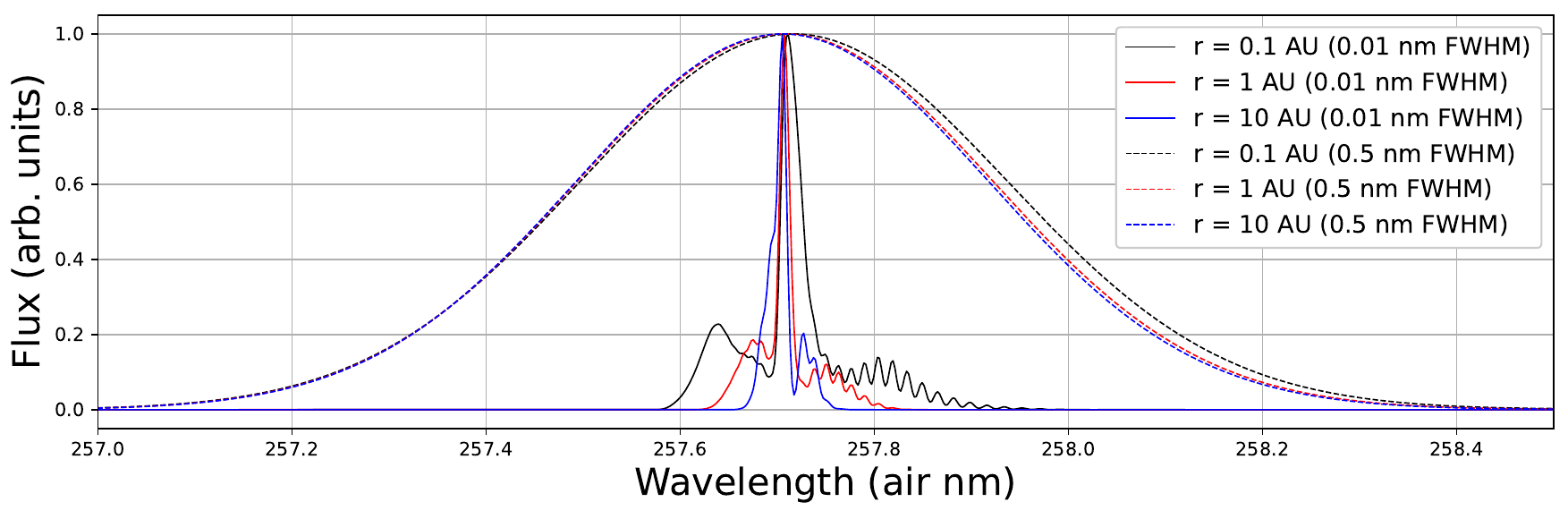}
    \caption{Synthetic fluorescence spectra of the CS ultraviolet $(0,0)$ band at $r_\textrm{h} = 0.1, 1, 10$ au at high (0.01 nm FWHM, solid lines) and low (0.5 nm FWHM, dashed lines) spectral resolution. The vertical axis is proportional to J s$^{-1}$ molecule$^{-1}$, with all models normalized to a maximum value of 1 to highlight changes to band shape.}\label{fig:distbands}
\end{figure*}

{Lastly, we investigated whether heliocentric distance plays a role in the distribution of features within a band. Such an effect may impact comparisons to high-resolution spectra.} As shown in Fig.~\ref{fig:distbands}, changes in heliocentric distance can have a modest impact on the distribution of lines within a band. However, at the low spectral resolution with which CS has historically been observed ($\sim$0.5\AA~FWHM), the effect on the shape of the bands is negligible. In short, re-scaling of CS band luminosities from 1~au by a factor of $1/r_\textrm{h}^2$ for the purposes of deriving column densities and/or abundances is justified.

\section{Astrophysical Implications}\label{sec:discussion}
The continued stability of the CS A$^1\Pi$-X$^1\Sigma$ fluorescence efficiencies, despite marked improvements in solar spectral references, an expanded rovibronic structure, and experimentally-benchmarked transition rates, confirms that the NUV-derived CS column densities in cometary comae from 1980 onward are accurate {given the present understanding of CS$_2$ as the CS parent molecule.} In turn, we are able to proceed in scrutinizing the cometary sulfur abundance catalog with renewed confidence in the measured column densities and production rates of CS, although the source of the molecule remains unclear \citep{Meier1997,roth2021leveraging,noonan2023evolution}. We note that our higher resolution calculations and ability to reliably calculate the Swings effect have implications for observations taken beyond the solar system, and in particular for protoplanetary disks. 

As stated earlier, CS has been observed in protoplanetary disks frequently in the {radio}, typically via the $J=2\rightarrow1$ rotational transition (see for example~\citealt{kama2019abundant}). As such, these observations probe the cold mid-plane of protoplanetary disks. NUV fluorescence of the CS molecule is extremely unlikely in this region due to the optically thick dust, preventing incident stellar UV radiation from exciting CS. Any CS fluorescence in the UV observed in protoplanetary disks would be linked to the upper layers of the disk, in the illuminated photodissociation region~\citep{Keyte2023azimuthal}. Despite many observations of protoplanetary disks encompassing the wavelength range of the CS A$^1\Pi$-X$^1\Sigma$ region as part of the ULYSSES survey\footnote{https://ullyses.stsci.edu/}~(e.g.,~\citealt{arulanantham2018uv,espaillat2022odysseus,skinner2022hst}) we were unable to find any studies that confirmed the presence of this band. With the improved models for the CS A$^1\Pi$-X$^1\Sigma$ $(0,0)$ band presented in this paper it may be possible to directly compare the abundances of CS in the midplane to CS that has been transported up into the PDR and is fluorescing. A similar strategy to capture the temperature structure in the disk has been employed by observing both the CO A$^1\Pi$-X$^1\Sigma^+$ band and the CO rovibrational lines at 4.7-5 $\mu$m \citep{arulanantham2018uv}. Detections of the CS A$^1\Pi$-X$^1\Sigma$ band {in conjunction with radio observations of the} rotational emissions would provide insight into the variability of the S/O abundance as a function of vertical location in the disk. A systematic re-analysis of NUV observations of T-Tauri stars obtained with space telescopes like HST and IUE could produce such results.

{With regard to comets, this work was initiated following the identification of discrepancies in sulfur abundances derived from {radio} and UV observations of comets, as discussed in \cite{noonan2023evolution}. A possible source of these discrepancies, inaccurate fluorescence efficiencies for the A$^1\Pi$-X$^1\Sigma$ bands of CS, has now been definitively ruled out. The present $g$-factors, utilizing an updated solar spectrum, expanded energy level structure, and experimentally-benchmarked transition rates, recovers comparable $g$-factors to previous studies of the molecule. In particular, we observe that the assumption of pure fluorescence, i.e. no collisional or time-dependent effects, as well as the possibility of a partially thermalized X$^1\Sigma$ ground state leads to negligible changes in the band luminosities. Time-dependent models also recover similar behaviors as seen in previous modeling efforts, such as the work of \cite{Swamy1993}, while maintaining band luminosities close to those previously reported. Though we have neglected collisional effects, their inclusion is expected to change the present $g$-factors for the NUV bands in a negligible way. These findings suggest that the NUV fluorescence properties, while contributing to minor changes in the derived CS abundances, are not the primary cause of the factor of 2 - 5 enhancement in CS abundances from NUV observations, leaving us to speculate on the source.}

{Historically, CS$_2$ has often been invoked as the parent of CS, given its short photodissociative lifetime under solar radiation and dominant dissociation channel that produces CS \citep{huebner2015photoionization}. Investigations of the spatial profile of the CS A$^1\Pi$-X$^1\Sigma$ emission have shown best fit lifetimes between 340-1000 seconds, all consistent with CS$_2$ as a parent molecule. However, observations in the radio have recently shown that the spatial distribution of the CS ($J$=7-6) transition is better characterized by an extended source with a scale length nearer 2000 km, rather than the 250-750 km expected for CS$_2$ \citep{roth2021leveraging}. If such an extended source is refractory in nature, this may be contributing relatively hot CS$_2$ to the coma - and as such may produce a unique signature for the CS produced from the dissociation. Such an effect may help to explain the broad CS A$^1\Pi$-X$^1\Sigma$ band seen in many comets that have had large apertures extracted in the NUV \citep{Meier1997}; one can imagine that a cooling CS population expanding from the nucleus will have its peak wavelength shift blueward as it expands while also losing the contrast of the P and R branches on the blue and red wings, respectively (as shown in Figure \ref{fig:bradfield_highres}). Such an effect may be difficult to resolve in the NUV, given the narrow slit width of the STIS instrument compared to the beam size of radio observatories. However, such an effect does not explain why CS appears to be more abundant in the NUV observations compared to the radio.}

{One possible source for this discrepancy could be the result of simple pointing geometry. NUV observations of comets with STIS are typically aligned with the slit perpendicular to the Sun-comet line to avoid orientation issues with HST (see Figure 1 of \cite{noonan2023evolution}). Radio observations obtained with relatively wide beam widths (~10-15'') cover a significant portion of the inner coma within the beam, while long slit spectroscopy can only capture a narrow slice (52$\times$0.2'' or 52$\times$0.5'' for STIS) at many different nucleocentric distances; radio observations are implicitly more resilient to asymmetric production from the surface due to beam size. If the HST STIS observations are biased in such a manner due to the typical slit orientation, and this unintentionally either increases the derived CS abundance or decreases the derived H$_2$O abundance for observations, that effect would be proportional to the geocentric distance of the object when observed. Comet C/1999 H1 (Lee) was observed with both STIS \edit{(NUV)} and \edit{IRAM+JCMT+CSO+SEST (radio)} in the summer of 1999, when it was 1.2 au from the Earth~\citep{Biver2000}. While no CS or H$_2$O production rates have been officially published for the STIS data, \cite{Biver2000} state that \cite{Feldman1999Lee} found a CS/H$_2$O value within 30\% of their own. This marks the only time that STIS and radio measurements of the CS/H$_2$O ratio have agreed closely in the last 25 years, with the notable exception of C/1995 O1 (Hale-Bopp)~ \citep{weaver1997,weaver1999,biver2002}, which was measured with both IUE and HST's FOS and STIS at large geocentric distances (3 - 5 au) pre- and post-perihelion. It is also notable that this agreement coincides with the second largest geocentric distance for a comet observed with both NUV and radio assets (C/2009 P1 (Garradd) was observed at $\Delta$=1.72 au). The largest disagreement comes from 46P/Wirtanen, which was observed at just 0.17 au from the Earth \citep{noonan2023evolution}, and had the lowest total production rate. A deeper investigation of these factors and their interplay is beyond the scope of this work, and will be explored in a future manuscript.}

%To further investigate a possible correlation between geocentric distance and variation between derived ratios we plot the discrepancies between comets observed between 1995 and 2021 with STIS and radio, each independently deriving their own CS and H$_2$O production rates.}

\begin{figure*}
    \centering
   \includegraphics[scale=0.55]{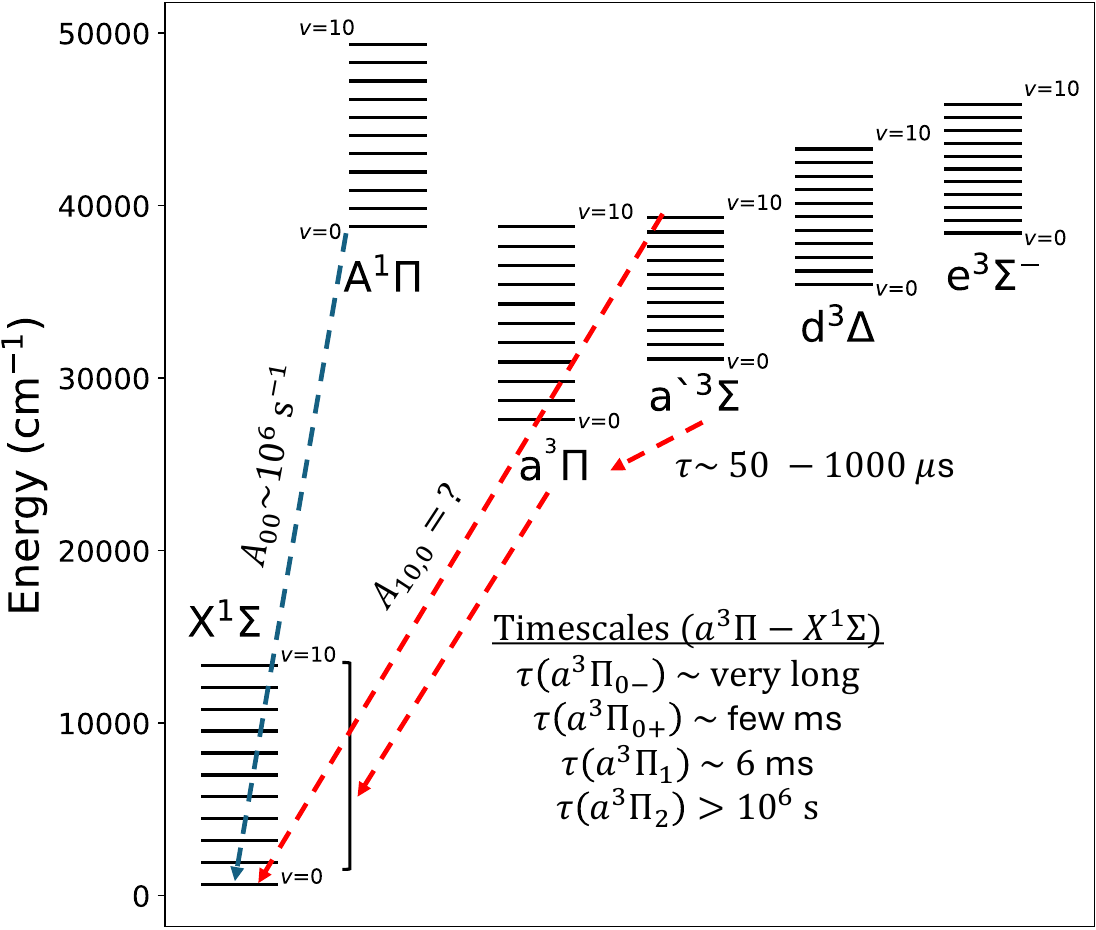}
    \caption{Vibronic energy levels $v = 0 - 10$ of two singlet and four triplet states of CS. Vibrational energies were computed from the published $T_e$, $\omega_e$, and $\omega_e{x_e}$ constants reported in \cite{Cossart1977} and \cite{Bergeman1981}. Transition rates ($A$) and radiative lifetimes ($\tau$) were taken from present work ($A_{00}$) and available literature~\citep{Li2013,Sun2020}. The unknown transition rate for the a$'^3\Sigma^+$-X$^1\Sigma~(10,0)$ band which overlaps the spectral extent of the A$^1\Pi$-X$^1\Sigma(0,0)$ band is indicated. All lifetimes are radiative, and are assumed to be summed over all relevant lower vibrational states.}\label{fig:energylevels}
\end{figure*}

{The molecular structure of CS and the details of its parent dissociation may allow an alternative explanation of the relative enhancement of UV emission (and thus CS abundance), as well as the extended nature of CS in the radio. Numerous experimental studies have resolved the molecular constants for the triplet states of CS, with the a$^3\Pi$, a$'^3\Sigma^+$, d$^3\Delta$, and e$^3\Sigma^-$ states lying between the X$^1\Sigma$ and A$^1\Pi$ states responsible for the NUV emission~\citep{Bergeman1981,Cossart1977}. \edit{Figure~\ref{fig:energylevels} shows the vibrational levels of these electronic states, along with radiative pathways relevant to the present discussion.} The energies of the A$^1\Pi$ vibrational levels are perturbed by interactions with the close-in-energy a$'^3\Sigma^+$ levels~\citep{Kuroko2022,Cossart1977}. In particular, the A$^1\Pi(v=0)$ level is strongly perturbed by the a$'^3\Sigma^+(v=10)$ levels, with substantial overlap between the transitions of the a$'^3\Sigma^+$-X$^1\Sigma~(10,0)$ and A$^1\Pi$-X$^1\Sigma$~(0,0) bands. Dissociation of the parent molecule(s) may populate the triplet states, which ultimately decay to a$^3\Pi$, located approximately 3~eV above X$^1\Sigma$. The a$^3\Pi$ state itself is split into four branches: a$^3\Pi_0^-$, a$^3\Pi_0^+$, a$^3\Pi_1$, and a$^3\Pi_2$. Both the a$^3\Pi_0^-$ and a$^3\Pi_2$ branches are expected to have `very long' lifetimes in excess of 10$^6$ seconds, with the other branches having radiative lifetimes much shorter than one second owing to spin-orbit coupling with the A$^1\Pi$ state~\citep{Moltzen1988,Li2013,Sun2020}.} 

{An alternative explanation for both the NUV and radio discrepancies may be realized as follows, assuming some production of CS within the triplet states instead of A$^1\Pi$ or X$^1\Sigma$. Assuming the levels of a$^3\Pi$ are statistically populated with respect to $\Omega=0,1,2$ following a dissociation event, $\sim$50\% of the a$^3\Pi$ population will be in either a$^3\Pi_0^-$ or a$^3\Pi_2$, with very long ($>10^6$~seconds) radiative lifetimes. \edit{Further, if the a$'^3\Sigma^+(v=10)$ level is strongly populated, prompt emission from the a$^3\Sigma^+$-X$^1\Sigma(10,0)$ band would contaminate A$^1\Pi$-X$^1\Sigma$~(0,0) band.} This effect would be difficult to distinguish from direct production in either singlet state \edit{at low spectral resolution.} Population remaining in the long-lived \edit{branches of the a$^3\Pi$ state} would not radiatively decay, but could be collisionally quenched. Taking the collision rate between CS and H$_2$O as $C = 8\times10^3/r^2$ with $r$ in kilometers for an H$_2$O production rate of $2\times10^{29}$~s$^{-1}$~\citep{Crovisier1983}, the resulting lifetime against quenching to the ground state of the triplet system at $r = 1000$~km is of order 125 seconds. For a lower water production rate of order $10^{28}$ molecules/sec, this increases to over 1000 seconds. Thus, long-lived triplet levels produced from the dissociation to CS would be initially invisible to radio observations, with potential increases in flux of the NUV $(0,0)$ band from overlapping prompt emission from the a$'^3\Sigma^+$-X$^1\Sigma~(10,0)$ band. Following collisional quenching of the long lived states to the singlet ground state, or the triplet ground state with a subsequent `fast' ($< 1$~second) radiative decay to the singlet system, the molecule would `turn-on' and begin emitting, mimicking an extended source in the radio. This scenario would explain both the extended nature of CS emission observed in radio~\citep{roth2021leveraging}, and if prompt emission is a sizable contributor to CS NUV emission, the relative enhancement of NUV-derived abundances as well.}

{The triplet states being responsible for both the extended nature of the radio emission, as well as enhancing the NUV band emission, is generally consistent with available literature. Notably, \cite{Butler1980} studied the dissociation fragments produced from photolysis of CS$_2$, with several bands between 370 and 610~nm likely attributed to progressions in the triplet systems. Photodissociation studies of CS$_2$ between 105 - 210~nm by \cite{Black1977} revealed near-unity production of triplet states following photodissociation from photons between $\sim125$~-~140~nm. While it is unclear which branches are populated, they predict lifetimes of a$^3\Pi$ less than 16 milliseconds, and noted that these measurements were weighted toward the $^3\Pi_0$ branches.  Quenching rate coefficients for CS a$^3\Pi$ were also measured by \cite{Black1977} for a number of gases (but not H$_2$O), which range between $10^{-11} - 10^{-15}$~cm$^3$~s$^{-1}$~molecule~$^{-1}$ at $T = 296$~K. Assuming nominal expansion velocities and water production rates and assuming a rate coefficient at the upper end of those measured by \cite{Butler1980}, the quenching timescale for CS a$^{3}\Pi$ is of order 100 - 1000 seconds. This is not definitive proof that the triplet states are  responsible for spatially-extended CS as inferred from radio observations, but is of the correct magnitude to motivate further exploration of this effect.}

{Given the low branching ratio ($<0.2\%$;~\citealt{huebner2015photoionization}) for producing triplet states of CS from CS$_2$, CS$_2$ alone can not explain the observed enhancement of the NUV-derived CS abundances, as well as the extended nature of CS emission in the radio. A parent with a longer lifetime than CS$_2$ would suggest a broader spatial profile of CS emission in the NUV that is inconsistent with observations.  From the above considerations, an additional parent with the following properties may explain all outstanding facets of this problem: (1) a short-lived parent molecule with a comparable photodissociation lifetime to CS$_2$, (2) said molecule must dissociative and strongly populate the excited states, both triplet and singlet, of CS.}

{We note that the current state-of-the-art photochemical lifetime data for CS is purely theoretical~(e.g.~\citealt{Heays2017}), and large errors in the photodissociation and/or photoionization lifetimes may lead to either enhanced or reduced CS abundances. Additional experiments and/or theoretical calculations of both dissociation properties of CS, as well as possible precursor molecules like H$_2$CS, would contribute to resolving the NUV/radio CS abundance discrepancy. In order to validate the assertion of triplet states producing the NUV/radio abundance discrepancy, such efforts should, where possible, probe the production of CS triplet states following dissociation from CS$_2$ and other potential precursors. Searches for the infrared emission of the a$'^3\Sigma^+$-a$^3\Pi$ bands of CS between approximately 2000 - 5000 cm$^{-1}$ would shed light on the extent of CS triplet state production from parent molecules in the coma. Accurate molecular constants necessary to compute the transition wavelengths for this purpose are available in the Tables of \cite{Krishnakumar2019}.}

{The current understanding of CS$_2$ alone, as of \cite{noonan2023evolution}, can not explain the abundance of CS inferred from UV observations when coupled with the results from \textit{Rosetta}~\citep{calmonte2016sulfur}. Moreover, the present calculations suggest that previously-calculated fluorescence efficiencies are not responsible for the relative enhancement of CS abundances derived from NUV observations. Prompt emission of CS, in either the A$^1\Pi$-X$^1\Sigma(0,0)$ or a$'^3\Sigma^+$-X$^1\Sigma~(10,0)$ bands following dissociation from parent species could contribute to NUV emission features, complicating an interpretation of the CS abundance.} Prompt emission also populates a different array of rovibrational states, with typically higher rotational states compared to solar fluorescence (see e.g. the case of OH in \citealt{Ahearn2015}). These highly-excited states evolve differently in time compared to molecules formed in the ground rovibrational state, which can present a substantial challenge for accurate spectral modeling {of high-resolution spectra. High-resolution UV comet spectra, capable of spectrally separating the transitions within the A$^1\Pi$-X$^1\Sigma$ $(0,0)$ band would directly test the possibility of prompt emission contamination within the spectral extent of the band. The necessary measurements} could theoretically be carried out using the G230H Echelle grating for the HST's STIS instrument (R$\sim$114,000). However, the low throughput and narrow slit size would require that the target be both unusually bright and have a well-characterized orbit to minimize acquisition errors. {This is further complicated by HST's transition to the reduced gyro mode in mid-2024, which severely limits tracking of moving targets in the inner solar system. The feasibility of such an observing program for a particularly bright comet in the near future should be evaluated while HST is still available.}

\section{Conclusions}
{To-date, CS abundances derived from NUV observations appear to be systematically elevated in comparison to near-contemporaneous radio observations. NUV observations suggest a very short-lived source consistent with CS$_2$ as a parent molecule, while radio observations suggest an extended source.} The historical record of CS A$^1\Pi$-X$^1\Sigma$ fluorescence efficiencies covers a range of reported values, with minimal description of the molecular and solar data used in their production. As such, the confidence in existing column densities and abundances of CS derived via NUV observations of the A$^1\Pi$-X$^1\Sigma$~$(0,0)$ band was unclear. We reported new fluorescence efficiency calculations of the CS radical utilizing a large rovibronic structure and a new set of experimentally benchmarked transition rates. The present $g$-factor for the A$^1\Pi$-X$^1\Sigma$ $(0,0)$ band at $\sim$2580~\AA, $5.07\times10^{-4}$ photons s$^{-1}$ molecule$^{-1}$, is consistent with previous literature and suggests changes in derived CS abundances of $\leq 16\%$. {The updated model can not explain the majority of the apparent enhancement of CS abundances derived from NUV observations.} 

Concluding, we find that the extent of the rovibronic structure utilized in the calculations has a negligible impact on the fluorescence efficiency of the strongest systems. We report, for the first time, explicit calculation of the Swings effect for CS, and find that the bands are weakly sensitive to the heliocentric velocity of the emitting environment. In addition, the band shape changes negligibly as a function  of $r_\textrm{h}$, and scaling of $g$-factors at 1~au to a given heliocentric distance via $1/r_\textrm{h}^2$ is valid. Thermalization of the X$^1\Sigma$ ground state leads to observable changes in band shape, as realized in archived IUE comet spectra. However, this effect changes band the luminosities by only a few percent compared to the pure-fluorescence models, and does not impact any previously derived CS abundances.

From time-dependent propagation of the fluorescence equations, we find that CS initially formed in the ground rovibrational state quickly achieves fluorescence equilibrium, suggesting that the use of $g$-factors computed with the assumption of fluorescence equilibrium has not introduced large errors in existing abundance derivations. However, time-dependent models with a `hot' distribution of CS, formed initially at $T = 1000$~K evolves much differently in time {and produces a different band shape.} Models incorporating collisional (density-dependent) effects may be necessary to fully explain the observed band shapes of CS in the NUV. Refined modeling may provide insights into the possible parents of CS through an understanding of the initial rovibrational states of CS populated by photodissociation of parent molecules and their time evolution in the coma. A possible observing scheme with HST is noted which could provide sufficient spectral resolution to investigate {the complexities of the NUV $(0,0)$ band shape, including possible contamination from prompt emission.}

{Based on our models, we} describe possible avenues for investigating sulfur reservoirs in protoplanetary disks via UV fluorescence of CS. Such analyses may shed light on the temperature structure in the disk via comparisons of archival {radio} and UV observations in the midplane and photodissociation regions. Lastly, {we identified several complications that inhibit the interpretation of CS emission in both past and present NUV observations of comets. These possibilities may be summarized as follows:
\begin{itemize}
    \item The enhancement in NUV-derived CS abundance appears to scale with decreasing geocentric/heliocentric distance. The largest disagreements between NUV- and radio-derived CS abundances are for comets with low geocentric distances. Unintentional observational biases, e.g. long-slit versus wide-beam fields of view, may also contribute to the apparent abundance discrepancies.
    \item Long-lived triplet states populated following dissociative production of CS may contribute to an extended source of CS emission. Reasonable estimations of the collisional quenching rates of CS produced in long-lived triplet states are the appropriate order of magnitude to mimic an extended source at radio wavelengths. Searches for infrared emission from the a$'^3\Sigma^+$-a$^3\Pi$ bands, which were recently characterized by \cite{Krishnakumar2019}, would probe the extent of triplet state production from parent molecule dissociation.
    \item Prompt emission from either the A$^1\Pi$-X$^1\Sigma~(0,0)$ or a$'^3\Sigma^+$-X$^1\Sigma~(10,0)$ bands could enhance the flux within the wavelength window of the A$^1\Pi$-X$^1\Sigma$ $(0,0)$ band and inflate NUV-derived CS abundances. Experimental and/or theoretical studies of prompt emission from potential parent molecules should be undertaken to probe the extent of this contamination. Moreover, the extent to which long-lived CS a$^3\Pi$ branches are populated in the dissociation of potential parent molecules should also be pursued.
\end{itemize}}

We hope that the calculations presented here will be used to further investigate the discrepancies between NUV- and {radio}-derived sulfur abundances in cometary observations. The fluorescence model, input files, fluorescence efficiencies, and related supplementary files are provided on the Zenodo service (\url{https://zenodo.org/records/13313004}).

\bibliography{Main_Elsevier_format}{}

\bibliographystyle{aasjournal}
% QED

\printcredits

\end{document}